\documentclass[fleqn,twoside]{article}
\usepackage[headings]{espcrc2}

\readRCS
$Id: espcrc2.tex,v 1.2 2004/02/24 11:22:11 spepping Exp $
\usepackage{graphicx}
\usepackage[figuresright]{rotating}

\newcommand{\AmS}{{\protect\the\textfont2
  A\kern-.1667em\lower.5ex\hbox{M}\kern-.125emS}}

\title{Contamination of short GRBs by giant magnetar flares: Significance 
of downward revision in distance to SGR 1806--20}

\author{Paul A. Crowther\address[Shef]{Dept of Physics \& Astronomy, University of
Sheffield, Sheffield, S3 7RH, UK}, 
        Joanne L. Bibby\addressmark[Shef], 
        James P. Furness\addressmark[Shef],
        J. Simon Clark\address{Dept of Physics \& Astronomy, The Open University,
Milton Keynes, MK7 6AA, UK}}

\runtitle{SGR1806--20}
\runauthor{Crowther et al.}

\begin{document}

\begin{abstract} We highlight how the downward revision in the distance to 
the star cluster associated with SGR 1806--20 by Bibby et al. (2008) reconciles 
the apparent low contamination of BATSE short GRBs by intense flares from 
extragalactic magnetars without recourse to modifying the frequency of one 
such flare per 30 years per Milky Way galaxy. We also discuss the variety 
in progenitor initial masses of magnetars based upon cluster ages, ranging 
from $\sim 50 M_{\odot}$ for SGR 1806--20 and AXP CXOU J164710.2-455216 in Westerlund 1 
to $\sim 17 M_{\odot}$ for SGR 1900+14 according to Davies et al. (2009) and 
presumably also 1E 1841--045 if it originated from one of the massive RSG 
clusters \#2 or \#3. \vspace{1pc} 
\end{abstract}

\maketitle

\section{INTRODUCTION}

It is well known that the initial, intense $\gamma$-ray spike from giant 
flares of extragalactic magnetars -- locally classified as either Soft 
Gamma Repeaters (SGRs) or Anomalous X-ray Pulsars (AXPs) at kpc distances 
-- could be mistaken for cosmological short Gamma Ray Bursts (GRBs) (Hurley et al. 2005). 
Typical peak outputs of magnetar flares lie in the range 10$^{41}$ erg\,s$^{-1}$ 
(SGR 1806--20 in Jan 1979) to 10$^{45}$ erg\,s$^{-1}$ (SGR 0525--66 in Mar 1979) 
would not be detected in galaxies lying at Mpc distances, but the peak 
luminosity of the intense flare from SGR 1806--20 in Dec 2004 was 2$\times 
10^{47} d_{15}$ erg\,s$^{-1}$ where the usually adopted distance of SGR 1806--20 
is $d_{15}$ = 15 kpc. BATSE, aboard the Compton Gamma-Ray Observatory, 
would have detected the initial $\sim$0.1\,s spike from SGR 1806--20 had 
it originated from a galaxy 30--50 $d_{15}$ Mpc away.

For an adopted rate of one such intense flare, per 30 yr, per Milky Way
galaxy, a total of 200--300 $d_{15}^{3} (\tau/30 {\rm yr})^{-1}$ flares 
would have been detected by BATSE over its 9.5 year lifetime, 
representing 
40--60\% $d_{15}^{3} (\tau/30 {\rm yr})^{-1}$ of all short GRBs detected by BATSE 
(Hurley et al. 2005; Lazzati et al. 2005). 
Consequently, one would expect 
a strong correlation between the position of BATSE short bursts and nearby 
star-forming galaxies. However, Tanvir et al. (2005) found that only 10\% 
of short GRBs originated from within 30 Mpc, while Lazzati et al. (2005) 
concluded that the contribution of giant magnetar flares to short GRB 
statistics was no greater than $\sim$4\%. Only a few candidate extragalactic 
giant flares have been proposed, such as GRB051103 in M81 (Ofek et al. 
2006), although such claims remain inconclusive (Hurley et al. 2010).
Apparently, the frequency of intense magnetar flares is much lower 
than one per 30 years, which remains a very credible possibility, 
 unless  the widely adopted 15 kpc distance to SGR 1806--20 is incorrect.

\begin{figure*}
\includegraphics[width=15cm]{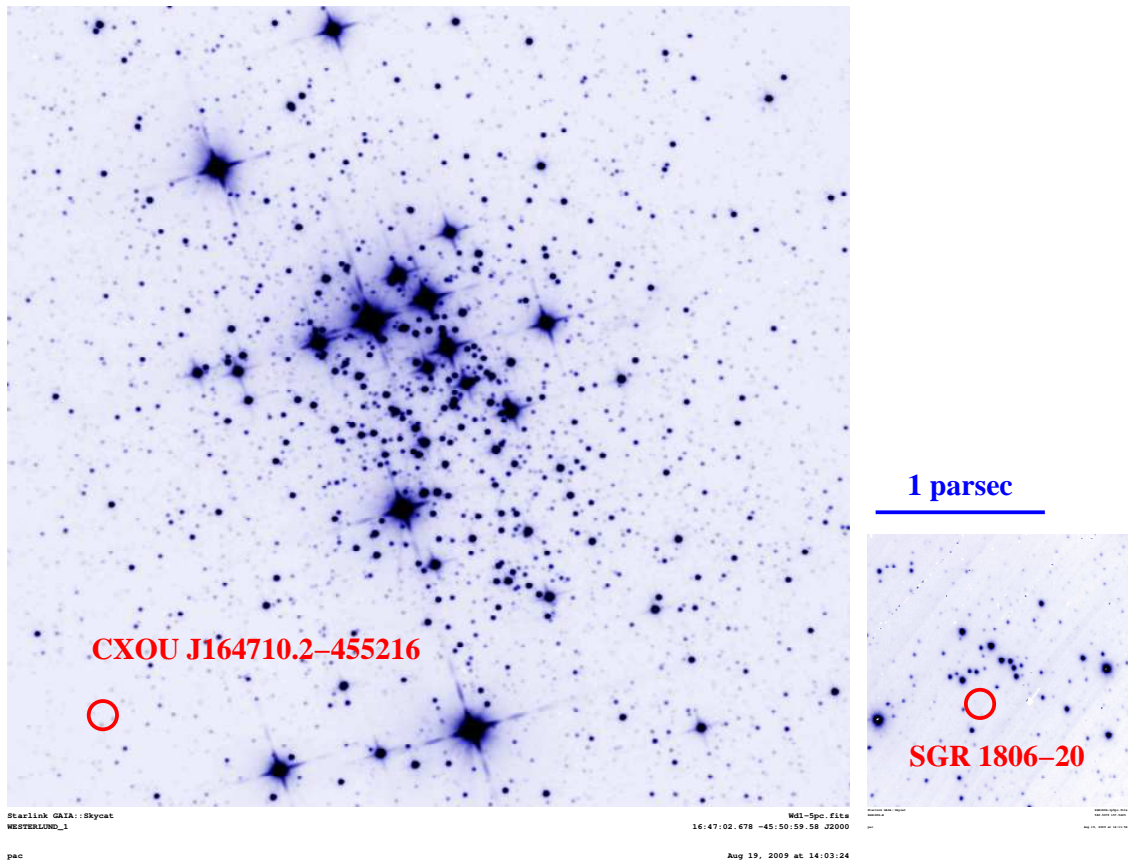}
\caption{Location of magnetars SGR~1806--20 and AXP CXOU J164710.2-455216 within visibly obscured young massive 
clusters: {\it Left} Westerlund 1 (NTT/SOFI, 200$\times$200 arcsec, 
equivalent 
to 
5$\times$5pc at $d$=5\,kpc); {\it right} Cl 1806--20 (Gemini-N/NIRI, 
34$\times$34  arcsec, equivalent to 1.5$\times$1.5 pc at $d$=8.7\,kpc).}
\label{fig0}
\end{figure*}

Fortunately, SGR 1806--20 lies towards a young star cluster, whose near-IR 
appearance is shown in Fig.~\ref{fig0}. Photometric (colour-magnitude 
diagrams) and spectroscopic (stellar subtypes) tools allow ages and 
main-sequence turn-off masses to be derived, albeit somewhat dependent 
upon isochrones from evolutionary models of massive stars. Walborn (2009) 
discusses the characteristics of the youngest star clusters, for which O 
stars represent the visually brightest stars up to ages of a few Myr 
(e.g. NGC~3603) with AF supergiants dominating for ages closer to 
4--10 Myr (e.g. Westerlund 1). Thereafter, the visual and near-IR 
appearance of star clusters is distinguished by red supergiants (e.g. RSG 
cluster \#1: Figer et al. 2006; RSG cluster \#2: Davies et al. 2007).

\section{DISTANCE TO SGR 1806--20}

The usual 15\,kpc distance towards SGR 1806--20 relies upon kinematics 
(Corbel et al. 1997; Eikenberry et al. 2004; Figer et al. 2004), but if we 
assume that it is physically associated with a visibly obscured ($A_{\rm 
K}$ = 3 mag), massive star cluster 
(Eikenberry et al. 2004; Figer et al. 2005), we may combine their 
physical properties with evolutionary models to obtain the age and 
(spectroscopic parallax) distance to the cluster. The likelihood of chance 
alignment between the cluster and magnetar is low. Indeed, Muno et al. 
(2006) obtained a confidence of $>$99.97\% for the association of magnetar 
 CXOU J164710.2--455217 (AXP) with the Westerlund 1 cluster (see Fig.~\ref{fig0}).
Both clusters host rare Luminous Blue Variables and Wolf-Rayet stars in 
addition to OB stars, although 
Westerlund~1 is an order of magnitude more massive (5--10$\times 10^{4} 
M_{\odot}$ Clark et al. 2005; Mengel \& Tacconi-Garman 2007; Brandner 
et al. 2008), so hosts many more evolved massive stars -- 24 WR 
stars 
(Crowther et al. 2006b) versus 4 WR stars in Cl 1806--20 (Eikenberry et 
al. 2004; Figer et al. 2005).

\begin{figure}[htbp]
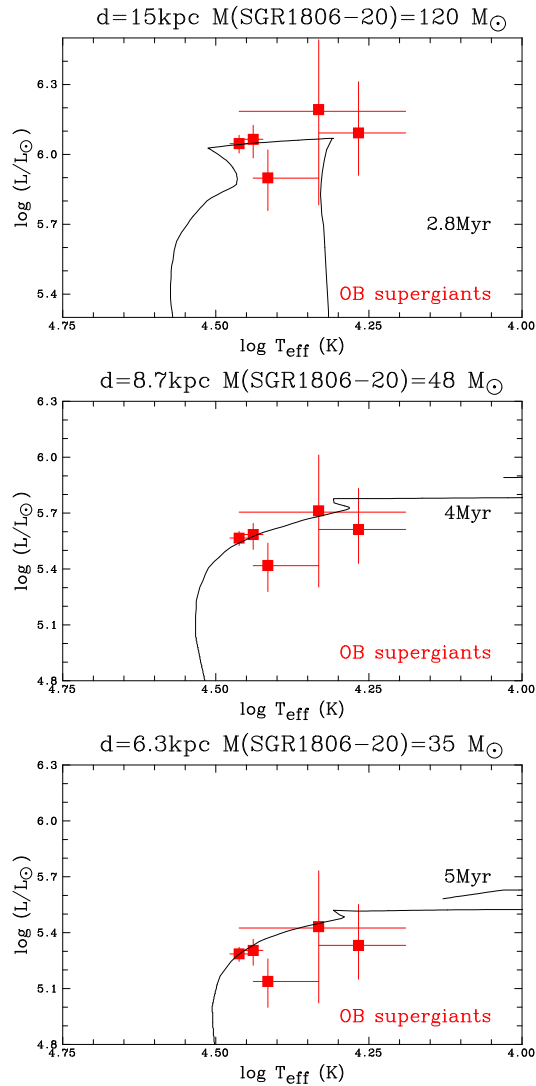

\includegraphics[width=4.8cm,angle=-90]{sgr1806_15_2p8.eps}
\includegraphics[width=4.8cm,angle=-90]{sgr1806_8p7_4.eps}
\includegraphics[width=4.8cm,angle=-90]{sgr1806_6p3_5.eps}
\caption{Meynet et al. (1994) isochrone fits (shown in black) to properties of B supergiants in SGR 1806--20 
(shown in red from Bibby et al. 2008) for ages of  2.8 Myr (top), 4 Myr 
(middle) and 5 Myr (bottom), from which distances of 15, 8.7 and 6.3 kpc 
are obtained. For the conventional 15 kpc distance to SGR 1806--20, the cluster OB stars 
would exhibit hypergiant spectral morphologies, while WC stars are not predicted
to exist at such young ages.}
\label{fig1}
\end{figure}

Bibby et al. (2008) presented Gemini-S/GNIRS near-IR spectroscopy of OB 
supergiants and non-dusty Wolf-Rayet stars in SGR 1806--20 from which a 
cluster distance of 8.7$^{+1.8}_{-1.5}$ kpc was obtained using a calibration of spectral 
type versus absolute magnitude. Reliable OB spectral types were obtained 
in two cases from comparison with the near-IR atlas of Hanson et al. 
(2005). Dust producing WR stars are highly unreliable as absolute 
magnitude calibrators since their near-IR appearance is dictated by the 
properties of their dust rather than underlying stellar continua (Crowther 
et al. 2006b). Main sequence OB stars would provide superior absolute 
magnitude calibrators with respect to Wolf-Rayet stars and OB supergiants, 
but these would require long near-IR integrations for reliable stellar 
classification even for present 8--10m telescopes.

\begin{figure*}[htbp]
\includegraphics[width=10cm]{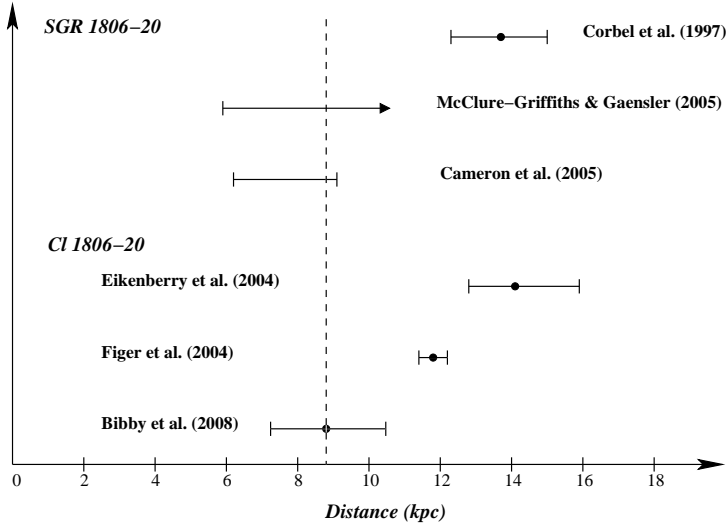}
\caption{Comparison between various magnetar and cluster distances to 
1806--20, uniformly adapted to a Galactic Centre distance of 8 kpc (Reid 1993).
Updated from Bibby et al. (2008).}
\label{fig2}
\end{figure*}

Alternatively, the simultaneous presence of WN and WC-type 
Wolf-Rayet stars in SGR 1806--20 suggest an age of 4$\pm$1 Myr, from which 
an independent distance measurement was obtained using isochrones for OB 
stars (Lejeune \& Schaerer 2001) based on the evolutionary models of 
Meynet et al. (1994) and the B supergiant temperature calibration of 
Crowther et al. (2006a). For ages of 2.8, 4 and 5 Myr, both cluster 
distances and magnetar masses were obtained, as presented in 
Fig~\ref{fig1}. Let us consider the properties of B supergiants
and WC stars for each of the three cases in turn,
\begin{itemize}
\item \underline{15 kpc} 
For this distance, the cluster B-type stars
studied by Bibby et al. (2008) would be exceptionally luminous, with
four of the five cases exceeding $\log (L/L_{\odot}) = 6$.
Instances of such luminous B supergiants are known, such as $\zeta^{1}$ Sco (Crowther
et al. 2006a). However, such stars display hypergiant spectral morphologies (B1.5\,Ia$^{+}$
for $\zeta^{1}$ Sco), whereas the B stars in Cl 1806--20 appear to be morphologically normal 
Ia, Iab or Ib supergiants. In addition, solar metallicity evolutionary models for 120 $M_{\odot}$ stars 
allowing for rotational mixing and contemporary mass-loss prescriptions
indicate that the WC phase is not predicted until after 3.5 Myr for a model initially
rotating at 40\% of critical (maximum) velocity (Hirschi, priv. comm.). For the 120  $M_{\odot}$ 
non-rotating case, the WC phase would be anticipated to {\it commence} after 2.8 Myr 
(Meynet \& Maeder 2003).
\item \underline{8.7 kpc}
For this distance, the cluster B-type stars
would be possess luminosities in the range $\log (L/L_{\odot})$ = 5.4 -- 5.7,
close to the $\log (L/L_{\odot})$ = 5.5 average of 25 O9.5--B3\,Ia supergiants studied by Crowther
et al. (2006a), the majority of which are morphologically normal. From Meynet \& Maeder (2003),
the hydrogen-burning phase of a non-rotating 50 $M_{\odot}$ star is predicted to last for 4.0 Myr, followed by 
short WN and WC phases, each of $\sim$0.1 Myr. In this case WC stars would be expected after $\sim$4.1 Myr.
Models initially rotating at 300 km\,s$^{-1}$ suggest longer main-sequence and WR phases, corresponding
to the onset of the WC phase after 5 Myr.

\item \underline{6.3 kpc}
For this distance, the cluster B-type stars
would be possess luminosities in the range $\log (L/L_{\odot})$ = 5.1 -- 5.5, somewhat below the
average of the Ia supergiants studied by Crowther et al. (2006a), and more representative of Iab or Ib
supergiants (Searle et al. 2008). Our GNIRS near-IR spectroscopic datasets does not allow us to discriminate
between Ia and Ib supergiants.  Again, turning to the WR population, from Meynet \& Maeder (2003), a
non-rotating 40 $M_{\odot}$ solar metallicity model would not be expected to remain as a WN
star prior to core-collapse, whereas a model initially rotating at 300 km\,s$^{-1}$ is predicted to 
advance to the WC phase only after $\sim$5.7 Myr.
\end{itemize}

\begin{table*}
\caption{Progenitor masses of magnetars based on associated star clusters.}\label{table1}
\begin{tabular}{l@{\hspace{2mm}}l@{\hspace{2mm}}l@{\hspace{2mm}}l@{\hspace{2mm}}l@{\hspace{2mm}}l}
\hline
SGR/AXP & Cluster & $d$   & Age  &  Mass  & Reference \\
        &         & kpc   & Myr       & $M_{\odot}$ & \\ 
\hline
1806--20 & Cl 1806--20 & 8.7$^{+1.8}_{-1.5}$ & 4$\pm$1 & 48$_{-8}^{+20}$ & Bibby et al. (2008)\\
CXOU J164710.2-455216 & Westerlund 1 & 5 & $\sim$4.5 & $\sim$55 & Clark et al. (2008)\\
1900+14 & Cl 1900+14 & 12.5  & 14$\pm$1 & 17$\pm$2 & Davies et al. 
(2009)\\
1E 1841--045 & RSGC \#2 or \#3 & 6 & 16$\pm$4 & $\sim$15 & Clark et al. 
(2009)\\
\hline
\end{tabular}
\end{table*}

Overall, a spectroscopic distance of 7--10 kpc is favoured from the simultaneous
presence of normal B  supergiants, plus WN and WC stars in Cl 1806--20. This cluster distance is compared to 
previous work in Fig.~\ref{fig2}, and is consistent with both the Cameron et al. (2005) 
magnetar distance and McClure-Griffiths \& Gaensler (2005) reanalysis thereof.

\section{CONTRIBUTION OF EXTRAGALACTIC GIANT FLARES TO BATSE SHORT BURSTS}

Overall, we find a reduced cluster distance of $d_{15} = 0.58\pm0.1$, and 
since the contribution of giant flares to BATSE short GRB statistics 
scales with $d_{15}^{3}$ we find that giant flares contribute 
8$_{-4}^{+5}$\% of all BATSE short bursts, for the {\it canonical} rate of 
one giant flare per 30 yr per Milky Way galaxy.  Of course, the
frequency of once per 30 years is merely a crude upper limit, since it
originates from the fact that solely one giant flare (from SGR
1806--20) has been detected
from a Milky Way source since the advent of gamma-ray telescopes. Nevertheless,
our result largely resolves the 
previous (low) rate problem, namely (i) $\sim$4\% from the scarcity of 
BATSE sources with spectral properties similar to giant flares (Lazzati et 
al. 2005); (ii) $\sim$10\% from giant flares within 30 Mpc (Tanvir et al. 
2005); (iii) $<15$\% from the lack of local host galaxies for several 
well-localized short GRBs (Nakar et al.  2006); (iv) a few percent 
contamination from  
the absence of giant flares from galaxies in the Virgo cluster (Popov \& 
Stern 2006).

\section{MAGNETAR MASSES FROM ASSOCIATED CLUSTERS}

As discussed above, associated clusters allow progenitor masses of 
magnetars to be estimated, since magnetars are believed to be young 
objects. For SGR 1806--20, Bibby et al. (2008) infer a progenitor mass of 
48$_{-8}^{+20} M_{\odot}$, adding to evidence linking some magnetars to 
very massive stars. Identical conclusions were reached by Muno et al. 
(2006) for the AXP CXOU J164710.2-455216  in the star cluster 
Westerlund~1 (Clark et al. 2005) whose age and stellar content is remarkably similar to Cl 
1806--20. Other observations have also been used to link magnetars to high
mass progenitors. For example, Gaensler et al. (2005) have interpreted a hydrogen
shell centred upon the AXP 1E 1048.1--5937
as a wind blown bubble from its 30--40 $M_{\odot}$ progenitor.

However, lower progenitor mass magnetars are also known. SGR 1900+14 
possesses a pair of M5 supergiants (Vrba et al. 1996; 2000). 
Davies et al. (2009) have recently presented deep K-band imaging
of these red supergiants, in which faint members of the associated
cluster Cl 1900+14 can be identified. Davies et al. (2009) have
obtained a kinematic distance  of 12.5 kpc to these M supergiants,
and an evolutionary age of 14$\pm$1 Myr, 
based upon contemporary solar metallicity evolutionary
models of Meynet \& Maeder (2000) plus bolometric corrections of
Levesque et al. (2005). If one assumes that the progenitor
of SGR 1900+14 originated from this cluster, an initial mass of
17$\pm$2 $M_{\odot}$ can be deduced.
Similar results  would be obtained for the AXP 1E 1841--045 (within Kes 73) 
{\it if} it  originated from either RSG cluster \#2 (Davies et al. 2007) 
or RSG cluster \#3 (Clark et al. 2009). These are located at a similar 
distance to 1E 1841--045, within a few degrees of its sight-line,
although the likelyhood of physical assocation is naturally much
weaker than for SGR 1806--20, CXOU J164710.2-455216 and SGR 1900+14.


Table~\ref{table1} provides a summary of magnetar progenitor masses 
inferred from potentially associated star clusters. In the Milky Way, it 
is apparent that magnetars can be produced by both $\sim 15 M_{\odot}$ and 
$\sim 50 M_{\odot}$ stars. The former presumably originated in Type II-P 
SN following the red supergiant (RSG) phase, while the latter are likely 
to have originated in a Type Ibc SN following the WR phase. From 
comparison with Heger et al. (2003), certainly SGR 1900+14 and possibly 
SGR 1806--20 would be expected to produce neutron star remnants at Solar 
metallicity. For metal poor populations the latter channel is unlikely to 
be available since reduced mass-loss rates during the main sequence and 
post-main sequence phases would likely lead to the formation of a black 
hole for such high initial mass stars. {No direct 
evidence for supernova remnants is known in Cl 1806--20, Westerlund~1 or 
Cl 1900+14. Presumably this is because the combined stellar winds from other
cluster members will blow a sufficiently large cavity that a supernova
remnant could expand into without showing any observable emission.}

Finally, it should be bourne in mind that cluster ages and magnetar 
progenitor masses rely upon the reliability of evolutionary models. Values 
quoted herein are based upon a set of assumptions which have subsequently 
been improved upon through mass-loss rate prescriptions and allowance for 
rotational mixing (Meynet \& Maeder 2000). Close binary evolution also 
further complicates inferred main-sequence turn-off masses. Therefore, one 
should not necessarily treat specific progenitor masses or cluster ages as 
robust, although the clear differences between the stellar content 
of clusters 1806--20 and 1900+14 undoubtedly demonstrate distinct channels 
leading to the production of magnetars.

\end{document}